\documentclass[sigconf,screen]{acmart}

\setcopyright{none}
\copyrightyear{2022}
\acmYear{2022}

\acmConference[Registered Report at ESEM'22]{Registered Report}{June 30, 2022}{Finland, FI}
\acmBooktitle{ACM/IEEE International Symposium on Empirical Software Engineering and Measurement (ESEM'22). available as ArXiv 2209.07211} 
\acmDOI{10.48550/arXiv.2209.07211}
\acmISBN{Available as ArXiv 2209.07211}

\usepackage{multicol}
\usepackage{color, colortbl}
\usepackage{todonotes}
\usepackage{multirow}
\usepackage{graphicx}
\usepackage{caption}
\usepackage{lineno}[color]

\definecolor{Gray}{gray}{0.9}

\long\def\longcaption#1#2{\centering\begin{minipage}{#1}\vspace{-0.7\baselineskip}\footnotesize\noindent\emph{#2}\end{minipage}}

\def\SAPR{APR4Sec}
\def\shorttitle{PPM}

\begin{document}

\title[Human acceptance of security patches suggested by APR tools]{On the acceptance by code reviewers of candidate security patches suggested by Automated Program Repair tools}
\subtitle{Registered Report of the Experimental Design}

\author{\href{https://orcid.org/0000-0003-3207-7662}{Aurora Papotti}}
\affiliation{%
  \institution{Vrije Universiteit Amsterdam}
  \city{Amsterdam}
  \country{The Netherlands}}
\email{a.papotti@vu.nl}
  
\author{\href{https://orcid.org/0000-0002-6682-4243}{Ranindya Paramitha}}
\affiliation{%
  \institution{University of Trento}
  \city{Trento}
  \country{Italy}}
\email{ranindya.paramitha@unitn.it}

\author{\href{https://orcid.org/0000-0002-1091-8486}{Fabio Massacci}}
  \affiliation{%
  \institution{University of Trento}
  \city{Trento}
  \country{Italy}}
\affiliation{%
   \institution{Vrije Universiteit Amsterdam}
  \city{Amsterdam}
  \country{Netherlands}}
\email{fabio.massacci@ieee.org}

\renewcommand{\shortauthors}{Papotti,  Paramitha, Massacci}

\begin{abstract}
\textbf{Background:} 
Testing and validation of the semantic correctness of patches provided by tools for Automated Program Repairs (APR) has received a lot of attention. Yet, the eventual acceptance or rejection of suggested patches for real world projects by humans patch reviewers has received a limited attention.\\
\textbf{Objective:} To address this issue, we plan to investigate whether (possibly incorrect) security patches suggested by APR tools
are recognized by human reviewers. We also want to investigate whether knowing that a patch was produced by an allegedly specialized tool does change the decision of human reviewers. \\
\textbf{Method:} In the first phase, using a balanced design, we propose to human 
reviewers a combination of patches proposed by APR tools for different 
vulnerabilities and ask reviewers to adopt or reject 
the proposed patches. In the second phase, we tell participants that some of the proposed 
patches were generated by security specialized tools (even if the tool was actually a 
`normal' APR tool) and measure whether the human reviewers 
would change their decision to adopt or reject a patch.\\
\textbf{Limitations:} The experiment will be conducted in an academic setting, and to maintain power, it will focus on a limited sample of popular APR tools and popular vulnerability types.
\end{abstract}


\maketitle

\pagestyle{plain}

\section{Introduction}
\label{sec:intro}

The current trend in software development pushes for shorter release cycles \cite{wang2020:updates:java} and several tools support 
developers in this process \cite{shahin2017continuous:review}. Unfortunately, the very 
process that generates quick updates and increase business opportunities is also a source 
of security vulnerabilities \cite{massacci2021technical:conference}. 
To address the trade-off between business opportunities and security risks \cite{massacci2021technical:magazine}, qualitative studies with developers have shown that 
developers would appreciate automated tools to quickly and 
correctly patch security vulnerabilities
\cite{pashchenko2020qualitative:dependency}.

In this respect, Automated Program Repair (APR) tools could be a promising avenue. The 
field is mature and the performance 
and correctness of the different tools have been the subject to significant automated 
benchmarking studies performed by independent researchers \cite{durieux2019empirical,kechagia2021evaluating,martinez2017automatic:assessment,chong2021assessing:student:code:review}. The applicability of APR tools designed for security has also been the subject of some comparative case studies~\cite{pinconschi2021comparative}. Recently, specialized tools for security vulnerabilities such as SeqTrans~\cite{chi2022seqtrans}, VuRLE~\cite{ma2017vurle}, and SEADER~\cite{zhang2021data} have also been recently proposed, albeit not yet independently evaluated. 

Yet, it has been shown that patches identified by APR tools may have passed all automatic 
tests and still be `semantically incorrect' in which they do not actually fix the bug or 
introduce bugs in other functionalities when subject to a manual validation \cite{liu2021critical:correcteness,wang2020automated:correcteness}. The percentages of 
`semantically incorrect' patches is significantly high even for mature and well studies 
tools: out of 395 bugs present in 
Defetcs4J, the best performing tool could semantically fix only 54 bugs, even when given 
the exact bug-fixing position  \cite[Table~3 vs Table~5]{liu2021critical:correcteness}.

Developers using APR tools might face candidate patches that are semantically incorrect: 
this is a \emph{change-based code review} problem
\cite{cohen2010modern,baum2016factors} where the only difference is that the 
patch comes from an APR tool rather than a human developer
as it happens in companies~\cite{sadowski2018modern} or open source 
projects~\cite{rigby2013convergent,rigby2014peer}. 

Hence, a partially unexplored overarching research 
question~\cite{zhang2022program:correcteness,cambronero2019characterizing, fry2012human} is 
to understand whether \emph{human code reviewers
will be able to discriminate between correct and incorrect security patches submitted by 
APR tools} (RQ1). 
Since different vulnerabilities require different patterns it might be that human patch 
reviewers would be able to discriminate more effectively for some vulnerabilities than 
others. There might be a difference also among tools using different patch strategies but 
to explore this option in details one would require that the number of patches generated by each tool is significantly larger than the one currently available \cite{liu2021critical:correcteness,wang2020automated:correcteness}.

Consequential to the research question above is understanding \emph{whether code reviewers decisions will be actually 
influenced by knowing that some patches come from a specialized security tool} (RQ2). There is already some evidence in 
the literature that providing this information can influence code reviewers. 
Such bias might not be necessarily for the good as APR tools designed for security (\SAPR) might not be necessarily 
more accurate than the general APR tools.

The purpose of this registered report is to present the design of an experiment with human code reviewers 
to understand the answers to the above two research directions.

\section{Related Work}
\label{sec:related-work}
\noindent \textbf{Bug repair.} Durieux et
al.~\cite{durieux2019empirical} performed a large-scale
evaluation of eleven APR tools on 2,141 general bugs from five
different benchmarks. The result of their experiment is not
promising as expected, the tools were able to fix only
0.7-9.9\% of the bugs. This study's outcome is affected by
patch overfitting in the Defects4J
dataset~\cite{just2014defects4j}. However, this study did not
assess the correctness of the patches, and they provide only a
comprehensive review of tools repairability. There are several studies~\cite{liu2019icseeval, kechagia2021evaluating, martinez2017automatic:assessment, ye2021automated:assessment, wang2020automated:correcteness} that assess APR patches correctness. These studies show that several APR generated patches
from APR tools are incorrect.

\noindent \textbf{Vulnerability repair.} Among the studies that
assess APR tools' effectiveness for security vulnerabilities,
there is the paper of Le Goues et al.~\cite{le2012systematic}.
This study evaluated GenProg (a generic APR tool enhanced for fixing
vulnerabilities) on the fixes for CVE-2011-1148. This paper focuses on the evaluation
of a single vulnerability, therefore, there is not a complete
evaluation of the tool's performances.   

Several tools have been implemented
specifically for vulnerability repairing. Abadi et
al.~\cite{abadi2011automatically} developed a tool specifically
for fixing injection vulnerabilities by placing sanitizers on
the right place in the code. More recently, 
SeqTrans~\cite{chi2022seqtrans}, VuRLE~\cite{ma2017vurle}, and 
SEADER~\cite{zhang2021data} are tools that leveraged existing 
repair patterns to produce vulnerability patches. Despite these studies, an independent evaluation of the the repair rate on a
broader set of vulnerabilities is still lacking as the the source code
of SeqTrans~\cite{chi2022seqtrans}, and SEADER~\cite{zhang2021data} has
been only recently published. 

Another study, regarding security vulnerabilities, performed by
Pinconschi et al.~\cite{pinconschi2021comparative}, compares ten
APR tools for C/C++ programs, using the concept of PoV (Proof
of Vulnerability) test. The aim of the test is to measure the
success rate of an APR technique. However, this study lacks of
semantic correctness assessments of the generated patches.

\noindent \textbf{APR vs. developers.}
Even though there have been a lot of studies in APR evaluation, these
machine-based evaluations have biases~\cite{liu2021critical:correcteness}. This fact
encourages human studies with APR tools~\cite{zhang2022program:correcteness, cambronero2019characterizing, fry2012human,tao2014automatically} to understand how effective actually
the patches in assisting development process. However, a study by Winter
et al.~\cite{winter2022let} found that this kind of human studies
is still rare, and make only 7\% in the Living Review~\cite{repair-living-review},
even less for patch adoption experiments. 

\vspace{0.5\baselineskip}
\noindent \fbox{\parbox{0.97\columnwidth}{\textbf{Key Novelty:} Building on these studies~\cite{liu2021critical:correcteness,zhang2022program:correcteness,cambronero2019characterizing,fry2012human,tao2014automatically,winter2022let}, we design an experiment with humans code reviewers to determine if Automated Program Repairs tools effectively support the identification of correct vulnerability fixes.}}
\vspace{0.5\baselineskip}

\noindent \textbf{Code Review and Secure Coding practices.}
Over the years several empirical studies on code inspections have been
conducted~\cite{kollanus2009survey}. Code reviewing occupies expensive
developer time, therefore, nowadays organizations are adopting modern
code review technique~\cite{cohen2010modern}.  Modern code review or
change-based review~\cite{baum2016factors}, is widely used across
companies~\cite{bacchelli2013expectations,sadowski2018modern}, and
community-driven projects~\cite{rigby2013convergent,rigby2014peer}. 

We investigated into studies which explicitly asked for implementing
secure coding practices. Naiakshina et al. ~\cite{naiakshina2017developers,naiakshina2018deception} conducted two
experiments with 40 computer science students, who have been
divided into two halves. The two groups received a different task
description. One description did not mention security, the other
one explicitly gave security instructions; As result, the group
without security instructions did not store passwords securely. 

There are few studies on checklists for contemporary review. Rong et
al.~\cite{rong2012effect}, through a study with students, found that
checklists were helping them during code review. In addition, Chong et
al. asserts that students were are able to anticipate potential defects and create a
relatively good code review checklist~\cite{chong2021assessing:student:code:review}. Finally,
a reports by Gonçalves et al.~\cite{gonccalves2020explicit}, explores
whether review checklists and guidance improve code review performances.

Braz et al.~\cite{braz2022less} explores both aspects of explicitly
asking for secure coding practices and providing checklists. She
conducted an online experiment with 150 developers, setting up three
different treatments. The baseline treatment consists in asking to the
participants to perform a code review without any special instructions.
In another treatment she explicitly asked to the developer to perform the
review from a security perspective. Finally, the third treatment
additionally asked developers to use a security checklist in their
review. The results showed that asking participants to focus on security
increases the probability of vulnerability detection, besides, the
presence does not significantly improve the outcome further. 

\vspace{0.5\baselineskip}
\noindent \fbox{\parbox{0.97\columnwidth}{\textbf{Key Novelty:} On the basis to these experiments~\cite{braz2022less,naiakshina2017developers,naiakshina2018deception}, we plan to randomly assign the participants to two different
treatments. One treatment is supposed to give to the participants a real
security information about the APR tools provided. Contrarily, we want to
provide a bogus information to the participants assigned to the other
treatment. }}
\vspace{0.5\baselineskip}

\section{Research questions}
\label{sec:rqs}
We structure our study around the two 
research questions: 

\vspace{0.5\baselineskip}
\noindent \fbox{\parbox{0.97\columnwidth}{\textbf{RQ1.} \textit{Will human code reviewers be able to discriminate between correct and incorrect security patches submitted by
the APR tools?}}}
\vspace{0.5\baselineskip}

As of today, all APR tools are 
research tools, with a great variety of user experience. Their `users', who 
are not the tool's 
inventors or competing researchers,
are essentially novices. From an 
internal validity perspective, this is 
an advantage, as our `users' know 
about the domain but not about the 
tool inner workings so they don't have 
a prejudiced prior belief on what the tool expected output should be.

We hypothesize that the number of wrong patches identified as
wrong, will be higher than the number of correct patches
identified as correct, and therefore adopted. 

\noindent\textbf{$H1.1$}: \textit{Wrong patches are more easily identified as wrong than correct patches are identified as correct.}

The practical impact of this hypothesis is that code reviews of APR patches 
is an effective last line of defense for gross mistakes. However, we further assume that is much harder to distinguish a partially correct patch
from a correct patch.

\noindent\textbf{$H1.2$}: \textit{Partially correct patches are equally identified as correct patches as actually correct patches.}

A further natural hypothesis is that specialized tools perform better than general purpose tools
and therefore an higher number of correct patches suggested by
security designed tools will be actually adopted by the code reviewers as they 
would more closely match what a security patch should be.

\noindent\textbf{$H1.3$}: \textit{Patches from APR tools designed for Security are adopted more often than
than patches suggested by generic APR tools.}

We have not specified whether such adoption happens \emph{irrespective of correctness} of the suggested patches. We suspect 
that correctness would not make a difference as the pattern of the patch rather than its actual semantic correctness would be 
a key measure of identification. This hypothesis leads to our second question.

\vspace{0.5\baselineskip}
\noindent \fbox{\parbox{0.97\columnwidth}{\textbf{RQ2.} \textit{Will code reviewers decisions be actually 
influenced by knowing that some patches come from a specialized security tool?}}}
\vspace{0.5\baselineskip}

To answer this question we need to perform a modest deception of participants that has been already used for password testing \cite{naiakshina2018deception}. We need to provide to a (random) subset of reviewer the true a bogus information that the one of the tool is a \SAPR when in reality it just a generic APR tool.
We formulate our corresponding hypothesis as follows:

\noindent\textbf{$H2.1$}: \emph{Both experimental and treatment groups will have same number of switches after revealing the security information.}

In other words, knowing that a tool is a security tool (even if is actually not such a 
tool) will create a bias into the decision making process of the code reviewer. Further, we 
hypothesize that after revealing the security information, the participants will tend to 
adopt the patches suggested by the security designed tools. Therefore, our second formal 
hypothesis is:

\noindent\textbf{$H2.2$}: \emph{The number of 
adopted patches from known security designed tools will be higher after the security 
information is revealed.}





\section{Experimental Protocol}
\label{sec:exp-protocol}

\begin{figure*}[t]
\centering


  \includegraphics[width=0.9\textwidth]{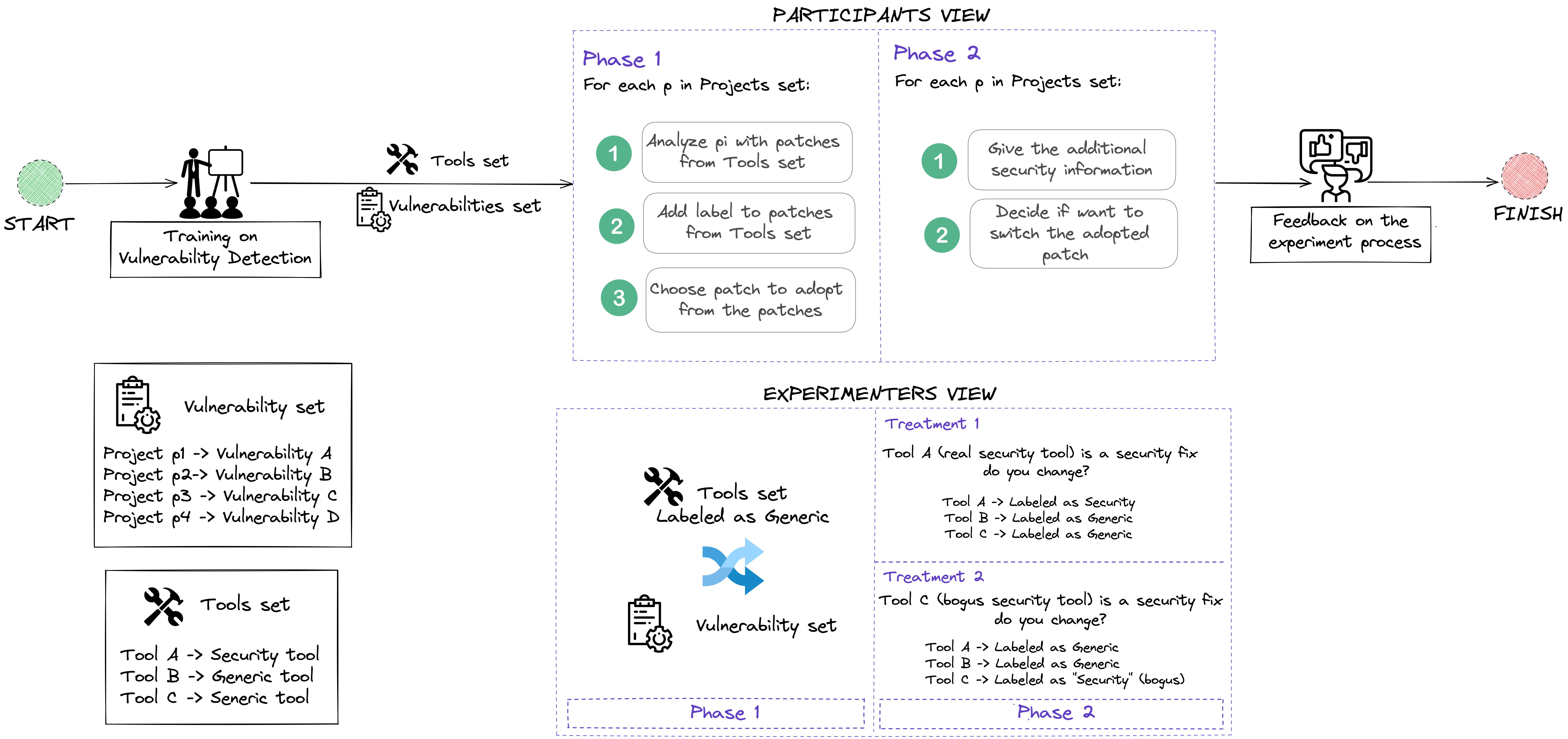}
  \caption{Steps of our experiment}\label{fig:experiment}
\end{figure*}

Figure~\ref{fig:experiment} summarizes the experimental protocol that we propose to address 
our research questions. We consider an APR tools set composed by a security tool A, and two 
generic tools B and C.
In the first phase of the experiments all the APR tools are labeled as generic. 
In the second phase of the experiment we plan to give to
one group the true security information, and to the other a bogus security information (the tools are labeled wrongly). We decided to set up different treatments on the basis of Braz et al.'s study~\cite{braz2022less}\footnote{We have already done a pilot, and the experimental plan in this work is adapted from the lesson-learned from the pilot.}.

\subsection{Execution Plan}
\label{subsec:exec-plan}
\noindent \textbf{Training.} The participants of the experiments have to complete two
training phases: \textit{(i)} Finding Vulnerabilities training, 
and \textit{(ii)} Fixing Vulnerabilities training.

The first preliminary activity aims to demonstrate that giving to developers a slice of the file, 
instead of the full file (during code review), leads to finding more vulnerabilities. 
Participants will therefore be trained on the identification of vulnerabilities into code
running for 1.5 hours two academic hours of 45 minutes. The slides give a general introduction about
vulnerabilities, and which one are the most common ones. Then, we provide
a more detailed explanation about injections, information disclosure, and
denial of service vulnerabilities, and how to recognize them in the code.
Then they will be asked to identify the vulnerabilities
in both a fragment of the code and in the full file. 

The duration of the training is an important experiment parameter. Comparing to other experimental activities this is considerably short: in the field of threat 
analysis and security requirements training sections last several 
hours~\cite{scandariato2015descriptive,wuyts2014empirical} or even days \cite{tuma2018two}.

In the domain considered for this experiment, Chong et al. \cite{chong2021assessing:student:code:review} performed the experiment 
after several weekly lectures with 60 minutes a week. On the opposite side of the spectrum, 
among the cited works, \cite{tao2014automatically} provides only 10 
minute tutorial. Several other works did some introduction or instruction or tutorial for 
their participants but they do not mention the length of the session \cite{naiakshina2017developers, naiakshina2018deception, naiakshina2018deception, cambronero2019characterizing, zhang2022program:correcteness, fry2012human}. Other works \cite{rong2012effect, gonccalves2020explicit} do not mention any training for their participants.  \cite{braz2022less} mentions training as one of their control variable but they did not actually conduct any training as their experiment was an online experiment.  

Eventually, we have opted for our construction by considering previous studies 
where professionals were involved and the minimum duration was indeed 1.5 hours \cite{allodi2020measuring,gramatica2015role}. This is essentially also a typical session 
of professional training session performed in industry\footnote{\url{https://www.secura.com/services/people/training-courses/secure-programming-training}}. Also the survey by Kollanus et al. \cite{kollanus2009survey} mentions the use of ``overview meetings" in most of the software inspection publications, which seems to imply that the duration was significant (as a 10 or 30 minutes presentation would hardly be considered a ``meeting'').

A second part of the training will happen a week later. The participants will receive a general introduction 
on how APR tools work with the IDE that will be used
to perform the code reviews of the patches suggested by the APR tools. 
We will provide to the participants the instructions to install an IDE plugin. 
Once the plugin is installed, the
vulnerable lines will be highlighted; then the students can choose which pair patch-tool to adopt.

Digital copies of lecture slides, 
technical documentation, etc. will be provided to
the participants and it can be
consulted at any time.

\noindent \textbf{Experiment - Phase 1} The experiment consists of three 
hours of physical laboratory. The participants will be separated
into different rooms according to their treatment (group) of
belonging to avoid spillover effects. In this phase, there are no
differences in the execution of the experiment between the two
treatments. Each room is supervised by 
an experimenter whose role does not go beyond the supervision of the
room and the technical support, s/he cannot
reply to questions regarding the solution of the correct
patches. 

The participants in both treatments will have to download a plugin in the Software Development IDE VsCode; 
The plugin will suggest different patches from different APR tools for each vulnerability. 
The participants will have several projects to evaluate: they will run each project in their own
environment, and use the plugin provided to analyze the highlighted vulnerability. 
Finally, they will select which patch to adopt to fix the vulnerability. 
On the Qualtrics submission tool, the participants will write, for each suggested patch, 
whether it is correct, partially correct, or wrong.

Once the participants have analyzed all projects, they can move to the second phase of the experiment. 

\noindent \textbf{Experiment - Phase 2}. In the second phase an
additional information about tools' type is given to the
participants. The students in the baseline treatment will receive a
correct information about the APR tools' type. Unlike, the students in
the second treatment will receive a bogus information. Once the students
receive the additional information, they can review their answers for
each project and decide whether to change their response, or keep their
choice made during the first phase of the
experiment. 

\noindent \textbf{Feedback} The participants will have to reply few questions regarding their perception of the task. The questions will be answered through an ordinal scale, moreover, we plan to design an open question for who wants to give additional comments, and give suggestions.

Since the experiment may be long, we plan breaks between the
experiment phases. Moreover, the participants can take breaks during
the execution of the experiment, or stop at any time they want, and
finish the experiment earlier.

\subsection{Measurement Plan}
\label{subsec:measurement}
The measurement plan describes where in the process we will collect the various 
variables that are further described in Section~\ref{subsec:vars}.

\noindent \textbf{Randomization.} We randomly separate the participants into two
groups (treatments), in the variable \emph{Assignment to a
Treatment}.

\noindent \textbf{Training.} The purpose of the training is to collect
information regarding the background of the participants as in the studies~\cite{braz2022less,naiakshina2017developers,naiakshina2018deception}. Knowing the
participant's background help us to better evaluate the results that we
collect in the experiment part.  
We collect the variables \emph{Knowledge of Security Vulnerabilities}, \emph{Knowledge of Java}, \emph{Knowledge of Software Development IDE} by
means of a Qualtrics questionnaire. For example we ask whether students
attended the tutorial, a full course,
have done vulnerability assessment in an internship or have true
professional assessment. 

We collect the \emph{Time spent on the video}, whether to check if the
students watched the video/material provided for the training part. 

\noindent \textbf{Experiment - Phase 1.} The purpose of the first phase
of the experiment is to determine if Automated Program Repairs tools
effectively support the developer in the identification of correct
vulnerability fixes. We collect the dependent variable to measure
correctness for proposed (tool,patch) whether \emph{Patch Classification}
adoption by mean of a questionnaire and a log as an additional measure, i.e. we ask whether the patch for CVE-2013-4379 proposed by Arja
is correct, partially correct, or wrong.

\noindent \textbf{Experiment - Phase 2.} The purpose of the second phase
of the experiment is to determine if \SAPR are better than generic APR tools. Also, we want to know if the `security'
information will lead to switches irrespective of correctness. We collect
the dependent variables \emph{Path-Adoption after security information by
pair (Tool, CVE)}, and \emph{Number of switches after security information by
pair (Tool, CVE)}.

\noindent \textbf{Feedback.} The experiment ends with few questions
regarding the perception of the tasks. We collect three variables: \emph{Process Understanding} to measure whether
the participants had a clear understanding of the experiment, therefore, of the tasks, \emph{Process Time} to check whether 
the participants had enough time to complete the experiment, \emph{Process Training} to control whether the training material was sufficient to carry out the tasks and on open ended \emph{Process Material} as a feedback if there was any other material that 
could have been useful or necessary to complete the tasks.

\section{Variables}
\label{subsec:vars}
Table~\ref{tab:vars} presents all the variables we consider in our experiment. 
\begin{table*}[]
\footnotesize
    \centering
    \caption{Experimental Variables}
    \label{tab:vars}
    \begin{tabular}{p{0.2\textwidth} p{0.58\textwidth} p{0.12\textwidth}}
        \toprule
         \textbf{Name} & \textbf{Description} & \textbf{Operationalize} \\
         \midrule
         \multicolumn{3}{l}{\textit{Independent variables (design)}} \\
         \midrule
         Assignment to a Treatment & Random assignments of participants to a treatment (true security information vs. bogus security information). & nominal (A) \\
        \midrule
         \multicolumn{3}{l}{\textit{Background variables}} \\
         \midrule
         Knowledge of Java & Self-reported concrete experience on Java & Ordinal scale (B) \\
         Knowledge of Security Vulnerabilities & Self-reported concrete experience on vulnerability assessment & Ordinal scale (B) \\
         Knowledge of SW Development IDE & Self-reported concrete experience on software development IDE & Ordinal scale  (B)\\
         \midrule
         \multicolumn{3}{l}{\textit{Dependent variables}} \\
         \midrule
         Patch-Classification by pair (Tool, CVE) &
         Classification of patch by the participants.
          &
          Ordinal scale (C)\\
         Patch-Adoption after security information by pair (Tool, CVE) &
         Classification of patch by the participants after receiving the security information.
          &
          Ordinal scale (C)\\
         Switches after security information &
         Number of patches classified differently by a participant before or after receiving information on the alleged security nature of tool.
          &
          ratio (A)\\
         \midrule
         \multicolumn{3}{l}{\textit{Experiment Validation Variables}} \\
         \midrule
         Time spent on training video & time (in minutes) to watch the video & ratio (A)\\
         Time spent on task & time (in minutes) to complete the task & ratio (A) \\
         Process Understanding & the participants had a clear understanding of the experiment & Ordinal scale (D)\\
         Process Time & the participants had enough time to complete the experiment & Ordinal scale (D) \\
         Process Training & the training material was sufficient to carry out the tasks & Ordinal scale (D) \\
         Process Material & material suggestions that can be useful to complete the tasks & Open Text \\
         Perceived Usefulness (PU) & self-reported usefulness of the prescribed technique & Ordinal scale (D)\\
        \bottomrule
    \end{tabular}
\begin{minipage}{0.9\textwidth}
\vspace{0.5\baselineskip}
\begin{description}
    \item[(A)] Automatically performed by the Qualtrics submission tool 
    \item[(B)] Multiple choice: no experience, attended a tutorial, attended a course, company internship, professional practice
    \item[(C)] Multiple choice: adopted, correct, partially correct, wrong
    \item[(D)] 5-point Likert scale: strongly disagree, disagree, neither-agree-nor-disagree, agree, strongly agree
\end{description}
\end{minipage}
    \end{table*}

\vspace{-\baselineskip}
\subsection*{Independent variables}
\textbf{Assignment to a Treatment.} For phase one, there will be diverse pairs of 
type of vulnerabilities and APR tool patches with different type of answers.
For phase two, this variable describes the random
assignment of each participant to one of the two treatments. For the
experiment, we designed two different treatments: \textit{(i)} in one
treatment a true security information about the nature of the tools is
given; \textit{(ii)} in the other a bogus security information about the nature of the tools is given.

\vspace*{-\baselineskip}
\subsection*{Background Variables}
The purpose of background variables is to ascertain if they have any experience in
different contexts (e.g. University projects,
personal developed projects, or professional
experience) that might impact the result of the experiment.\\
\textbf{Knowledge of Java.} Java experience of the participants. \\
\textbf{Knowledge of Security Vulnerabilities.} Vulnerability assessment experience of the participants. \\
\textbf{Knowledge of Software Development IDE.} Software Development IDE experience of the participants.


\subsection*{Dependent Variables}
\textbf{Patch-Classification by pair (Tool, CVE).} This variable
describes how participants classify the different patches proposed by the
selected tools for the experiment. It is used to answer to \textbf{RQ1}.\\
We designed the next two variables to answer \textbf{RQ2}.\\ 
\textbf{Patch-Adoption after security information by pair
(Tool, CVE).} This variable describes what happens once the participants receive the security information. \\
\textbf{Number of switches after security information by pair
(Tool, CVE).} This variable is 1 if a user has changed his/her decision to adopt a patch after receiving the information of which patches have been suggested by a security tool and 0 if no change has been made.

\subsection*{Experiment Validation variables}
We collect these variables in order to verify the correctness
of the experimental process.\\
\textbf{Time spent on training video} With this variable we measure the
time spent from each participant on the training video. Also, we verify
how many participants actually completed the training. \\
\textbf{Time spent on task} We designed this variable to measure the time
spent from each participant to classify the patches proposed
for each project. We measure the time by how long the participants spend on each page in the online form assessment. We do not envisage to measure time (beside the previously mentioned timer as overall interval) as this is not a fully controllable variable. We realistically assume that in 20 minutes per vulnerability one should have made a choice. \\
\textbf{Process Understanding} We ask to the participants if they had a
clear understanding of the experiment. We designed this variable to
measure the overall clearness of the experiment, and therefore what is
necessary to improve for the future. \\ 
\textbf{Process Time} We ask to the participants if the time provided to
complete the experiment is enough. \\  
\textbf{Process Training} We ask to the participants if the training is
sufficient to complete the experiment. We measure this variable to
collect feedback for future improvements of the experiment. \\
\textbf{Process Material} We measure if the provided material is useful
to carry out the tasks.\\
\textbf{Perceived usefulness (PU)} We ask to the participants to evaluate
the usefulness of the experiment.

\section{Ground Truth Dataset} 
We built a dataset that we intend to use for the execution of the
experiment. The dataset contains a set of vulnerabilities and APR tools that have been selected according to two criterion.

\noindent\textbf{Java Program Repairs}. A study on the repair of C++ vulnerabilities already exists~\cite{pinconschi2021comparative}. We want to focus on Java given the recent studies on normal fixes on Java \cite{durieux2019empirical}.

\noindent \textbf{Test-based}. We want to check the performance in presence of test cases present in industrial projects (Maven/Gradle).

\subsection{Choice of Vulnerabilities}
\label{subsec:choice-vuln}
We use the \textbf{Vul4J dataset} \cite{bui2022vul4j} as it is the only existing benchmark that satisfies our requirements. The dataset contains \textbf{79 vulnerabilities} from 51 real-world open-source Java projects (libraries, web frameworks, data-processing desktop apps, and CI/CD servers); it is extracted from the `Project KB' knowledge base~\cite{ponta2019manually}. 

We ran the tools on all 79 vulnerabilities in 
the dataset. However, only 14 of them 
actually generated patch/es from the tools. 
From this 14 vulnerabilities, we chose 7 in 
Table~\ref{tab:vulns} because \textit{(i)} 
they cover different vulnerability classes 
and \textit{(ii)} they have more than two 
(successfully-tested) patches from the 
selected tools.
    
\begin{table}
\caption{Vulnerabilities used in experiment.}
\label{tab:vulns}
\longcaption{\columnwidth}{}
\resizebox{\columnwidth}{!}{
\begin{tabular}{l|c|l|l}
\toprule
CVE & CWE & Keyword & Project \\ 
\midrule
CVE-2013-4378 & 79 & XSS & javamelody/javamelody \\ \hline
CVE-2016-9878 & 22 & Path Trav.& spring-projects/spring-framework \\ \hline
CVE-2018-1192 & 200 & Info. Disc. & cloudfoundry/uaa \\ \hline
CVE-2018-1324 & & & apache/commons-compress \\
CVE-2018-17202 &  &  & apache/commons-imaging \\
CVE-2018-1000864 & \multirow{-3}{*}{835} & \multirow{-3}{*}{Inf. Loop} & jenkinsci/jenkins \\ \hline
CVE-2019-10173 & 502 & Data deserialz. & x-stream/xstream \\ 
\bottomrule
\end{tabular}}
\end{table}

\subsection{Choice of Tools}
\label{subsec:choice-tools}
We limit the tools scope to Java test-based APR tools for which we have sufficient in-house expertise to validate the correctness of the APR generated patches for the experiment.
We selected both generic APR tools and, using four criterion. 

\noindent \textbf{Accessible Source Code}. We want to eliminate some elements of uncertainty in the vulnerability identification, and want to use 
Maven testing pipeline.

\noindent \textbf{Executable}. We do not intend to use tools that are no longer maintained, or not executable due to technical issues. 

\noindent \textbf{Extensible to any dataset}. We prefer to use tools that do
not require major efforts to make it work with a different dataset than
the tool's own recommended dataset.

\noindent\textbf{Generating (security) patches}. Each 
tool should possibly generate some patch for a vulnerability, even if the patch is not semantically correct.

Table~\ref{tab:tools} shows the APR tools that we selected. The applicability of the Maven test suite criteria is important for realism. We must be able to say to the participants that \emph{the selected patches passed all the tests}. In this way the participants will be put in the frame of mind that the eventual acceptance of the patch only depends on their code review.

The source code for the two \SAPR tools 
SEADER~\cite{zhang2021data} and
SeqTrans~\cite{chi2022seqtrans} have just been released. Therefore, we did
not have yet the chance to fully analyze them. We intend to use at least one of 
them for the experiment. 

We have considered 
the list of tools in the living review of Monperrus 
\cite{repair-living-review} and analyzed the relevant 
literature on APR tools and \SAPR. We chose the tools 
described in Table~\ref{tab:tools} as we mentioned in 
\ref{subsec:choice-tools}. We also limit the tool 
scope to Java test-based APR tools which then we can 
validate the correctness before using the patches for 
the experiment. We will add the APR tools designed for 
security (either SeqTrans~\cite{chi2022seqtrans} or 
SEADER~\cite{zhang2021data}) to Table~\ref{tab:latin-square} after we get their correctness result on the 
chosen CVEs.

\begin{table}
\footnotesize
\caption{Tools used in experiment.}
\label{tab:tools}
\longcaption{\columnwidth}{}
\begin{tabular}{l|c|c}
\toprule
Tool & Platform & Type \\ 
\midrule
ARJA~\cite{yuan2018arja} & \cellcolor{gray!20} &   \\
RSRepair-A~\cite{yuan2018arja} & \multirow{-2}{*}{\cellcolor{gray!20}
 Arja} & \\ \cline{1-2}
Cardumen~\cite{martinez2018ultra} &  &  \\ 
jGenProg~\cite{martinez2016astor} &  & \\
jKali~\cite{martinez2016astor} & \multirow{-3}{*}{Astor} &  \\ \cline{1-2}
TBar~\cite{liu2019tbar} & \cellcolor{gray!20}
Independent & \multirow{-6}{*}{Generic} \\ \hline
SEADER~\cite{zhang2021data} &  & \\ 
SeqTrans~\cite{chi2022seqtrans} & \multirow{-2}{*}{Independent} & \multirow{-2}{*}{Security} \\ 
\bottomrule
\end{tabular}
\end{table}

\subsection{Applying Tools on Vulnerabilities}
\label{subsec:pairing}
We have already started a preliminary analysis of the tools and the
vulnerabilities that we intend to use. Table~\ref{tab:latin-square} describes the patches classification of the APR tools for each vulnerability (from the dataset experiment) made by the experimenters. A cell of the table can assume three different values.
Each participant will be exposed to the same set of outputs corresponding to the patches of the vulnerabilities reported in Table~\ref{tab:latin-square}. They will receive six files and when opening a file they will receive a warning in Visual Studio that a vulnerability has been found and they will have to choose a fix identified by some APR tool. For each vulnerability \emph{all} generated patched will be shown. This design choice was made because some tools did \emph{not} produce any patch for several vulnerabilities. Therefore, we do not have enough patches per tool to run an experiment where we expose participants to different tools. Since the participants are exposed to the same set of outputs, there is no real `randomization'. However, the generation of patches was attempted from a wider sample of APR tools and some tools failed to report a patch while other succeeded (albeit with possibly a wrong or partially incorrect patches). We can consider this process a sufficient proxy for a randomising behavior (given the low success rate). We have considered the option of providing manually designed wrong patches but discarded it since they would not be generated by an APR tool and would not answer our research question as formulated.

\begin{table}
\caption{Preliminary Analysis of Possible Tools} 
\label{tab:latin-square}
\longcaption{\columnwidth}{Card.: Cardumen. jGen: jGenProg. APR tools designed for security} will be added by the time the experiment is carried out
\resizebox{\columnwidth}{!}{
\begin{tabular}{l|c|c|c|c|c|c}
\toprule
CVE & ARJA & Card. & jGen. & jKali & RSRepair & TBar \\
\midrule
CVE-2013-4378 & \cellcolor{yellow!20} PC &  &  &  & \cellcolor{yellow!20} PC & \cellcolor{yellow!20} PC \\
CVE-2016-9878 &  & \cellcolor{red!20} W & \cellcolor{red!20} W & \cellcolor{red!20} W &  & \cellcolor{red!20} W \\
CVE-2018-1192 & \cellcolor{green!20} C & \cellcolor{red!20} W & \cellcolor{yellow!20} PC &  &  & \cellcolor{green!20} C \\
CVE-2018-1324 & \cellcolor{green!20} C &  &  & \cellcolor{green!20} C & \cellcolor{green!20} C & \cellcolor{red!20} W \\
CVE-2018-17202 & \cellcolor{yellow!20} PC & \cellcolor{yellow!20} PC &  & \cellcolor{yellow!20} PC & \cellcolor{yellow!20} PC &  \\
CVE-2018-1000864 &  &  & \cellcolor{red!20} W & \cellcolor{red!20} W & \cellcolor{red!20} W & \cellcolor{red!20} W \\
CVE-2019-10173 & \cellcolor{yellow!20} PC & \cellcolor{red!20} W & \cellcolor{green!20} C & \cellcolor{yellow!20} PC &  &  \\
\bottomrule
\end{tabular}}
\end{table}

\noindent \textbf{C = Correct.} The experiment organizers considered the vulnerability patch correct.

\noindent \textbf{PC = Partially Correct.} The vulnerability patch fixes the vulnerability but might
introduce other functional errors as it changes the semantics of the execution in
some non-trivial ways (in the non-vulnerable case).

\noindent \textbf{W = Wrong.} The vulnerability patch is just plausible, but it is wrong.  It might have
passed all regression tests and the specific test showing that the vulnerability is present, but
according to the experimenters, it is clearly a wrong patch as it changes the semantics of the execution in a drastically different way.

If a cell is empty, does not mean that we did not select the tool, it means that there is no patch suggested by the tool for that vulnerability. Note that Table~\ref{tab:latin-square} does not report the classification for the tools SEADER~\cite{zhang2021data} and SeqTrans~\cite{chi2022seqtrans} because the source code has just been published. Thus, we did not have the chance yet to classify them; We plan to perform a further analysis, and include these security designed tools in the APR tools set for the experiment. 

For the very same lack of maturity in users' interfaces, participants would not actually \emph{use} the tools. They will receive the code 
changes recommended by the tool in a standardized and well known interface such as Visual Studio.

\section{Participants}
\label{sec:participants}
Our population is Master Computer Science students, with some differences
in the elective courses and program choices. All the participants are
students enrolled in the course Security Experiments and Measurements from
VU Amsterdam. The course is taught by the experimenters. 
We decided to perform the experiment with students as in the studies ~\cite{naiakshina2017developers,naiakshina2018deception,rong2012effect,chong2021assessing:student:code:review}.

The experiment will be performed during class time of the course,
and the purpose of this research methodology course is also to
introduce the students to the critical issues behind design,
execution, and measurements of security experiments.

In terms of learning outcome, through this experiment, the
students have the possibility to critical review the results of
the experiment and evaluate its statistical and practical
significance. In the course we do not evaluate the number of correct
responses given, but the student's capability to review and analyze 
the experiment results that we obtained. 

As we plan to have students as participants, this may affect
the outcome of the experiment. However, we believe that master
students have enough experience to participate into the
experiment, and we think that we can collect interesting
results and insights that can help us with the future research.
Moreover, obtaining significant results with students may
suggest that we designed a relevant experiment that we can
carry out with developers from companies in the future. 

\section{Analysis plan}
\label{subsec:analysis}
\textbf{Data cleaning.} We plan to perform a preliminary check on the collected data. All submissions without an explicit consent by the participants will be removed. Moreover, we will remove clearly invalid submission attempts if any, as measured by the process metrics.

\noindent\textbf{Ground Truth.} For the previous experiment we have
manually evaluated all patches generated by the tools in
advance and we compared them with the results of the
participants. We plan to follow the same procedure for this
experiment, and we plan to determine the correct number of
results, measuring the true positive, and false positive rates.

\noindent\textbf{Statistical Tests.} As we think we will not obtain distributed samples, we plan to use a non-parametric, Mann-Whitney test. Some of our hypothesis are about equivalence of treatments. To answer them, we will use TOST as a test of equivalence which was initially proposed by \cite{schuirmann1981hypothesis} and is widely used in pharmacological and food sciences to check whether the two treatments are equivalent within a specified range $\delta$~\cite{food2001guidance,meyners2012equivalence}. The underlying 
directional test will be again the Mann-Whitney test. In case we have too many zero values (i.e.\ many participants failed to recognize even \emph{some} lines of code) we will investigate the use of the combined test proposed by Lachenbruch 
\cite{lachenbruch2002analysis}.

\noindent\textbf{Validity threats.} We acknowledge that students'
background, knowledge, and practice may impact the experiment's
results. However, as mentioned in Section~\ref{sec:related-work}, several studies have been performed with students~\cite{naiakshina2017developers,naiakshina2018deception,rong2012effect,chong2021assessing:student:code:review}. Moreover, Salman et al~\cite{salman2015students} shows a comparison between students and professionals to understand how well students represent professionals as experimental subjects in SE research. The results show that both subject groups perform similarly when they apply a new approach for the first time. We also acknowledge that the time measurement would not be exactly
reflect the actual time the participants spend on the tasks. We plan to investigate these limitations with further studies.
However, we believe that we can still get significance results,
that will give us some strong basis to explore further in
the future, and replicate the experiment in different contexts;
such as with developers from companies. We also acknowledge that our sample is not representative of all
developers since we are considering only master students from a single
course. Therefore, to consider our study extensible and generalizable,
more studies should be designed and run.

\section*{Acknowledgments}
This work has been partly supported by 
the European Union H2020 Program under the Grant 952647 (AssureMOSS - \url{www.assuremoss.eu}). We would like to thank Quang-Cuong Bui, Duc-Ly Vu, and Riccardo Scandariato for many useful discussions on APR tools and their assessment and Ákos Milánkovich for providing the test Visual Studio plug-ins we have used in the pilot.
\subsection*{CRediT statements}
	\emph{Conceptualization:}	AP, RP, FM; 
	\emph{Methodology:} FM, AP, RP; 	
	\emph{Software:} not yet; 	
	\emph{Validation:} not yet;	
    \emph{Formal analysis:} not yet;	
    \emph{Investigation:} AP, RP;	
    \emph{Resources:} not yet;	
    \emph{Data Curation:} RP (Vulnerability tests in Table \ref{tab:latin-square}); 	
    \emph{Writing - Original Draft:} AP, RP
    \emph{Writing - Review \& Editing:} AP, RP, FM	
    \emph{Visualization:} AP	
    \emph{Supervision:} FM 
    \emph{Project administration:} FM	
    \emph{Funding acquisition:} FM	

\nolinenumbers
\bibliographystyle{ACM-Reference-Format}
\bibliography{short-names,literature}


\begin{thebibliography}{54}


\ifx \showCODEN    \undefined \def \showCODEN     #1{\unskip}     \fi
\ifx \showDOI      \undefined \def \showDOI       #1{#1}\fi
\ifx \showISBNx    \undefined \def \showISBNx     #1{\unskip}     \fi
\ifx \showISBNxiii \undefined \def \showISBNxiii  #1{\unskip}     \fi
\ifx \showISSN     \undefined \def \showISSN      #1{\unskip}     \fi
\ifx \showLCCN     \undefined \def \showLCCN      #1{\unskip}     \fi
\ifx \shownote     \undefined \def \shownote      #1{#1}          \fi
\ifx \showarticletitle \undefined \def \showarticletitle #1{#1}   \fi
\ifx \showURL      \undefined \def \showURL       {\relax}        \fi
\providecommand\bibfield[2]{#2}
\providecommand\bibinfo[2]{#2}
\providecommand\natexlab[1]{#1}
\providecommand\showeprint[2][]{arXiv:#2}

\bibitem[Abadi et~al\mbox{.}(2011)]%
        {abadi2011automatically}
\bibfield{author}{\bibinfo{person}{Aharon Abadi}, \bibinfo{person}{Ran
  Ettinger}, \bibinfo{person}{Yishai~A Feldman}, {and} \bibinfo{person}{Mati
  Shomrat}.} \bibinfo{year}{2011}\natexlab{}.
\newblock \showarticletitle{Automatically fixing security vulnerabilities in
  Java code}. In \bibinfo{booktitle}{\emph{Proc.\ OOPSLA'11}}.
  \bibinfo{pages}{3--4}.
\newblock


\bibitem[Allodi et~al\mbox{.}(2020)]%
        {allodi2020measuring}
\bibfield{author}{\bibinfo{person}{Luca Allodi}, \bibinfo{person}{Marco
  Cremonini}, \bibinfo{person}{Fabio Massacci}, {and} \bibinfo{person}{Woohyun
  Shim}.} \bibinfo{year}{2020}\natexlab{}.
\newblock \showarticletitle{Measuring the accuracy of software vulnerability
  assessments: experiments with students and professionals}.
\newblock \bibinfo{journal}{\emph{Empir. Softw. Eng.}} \bibinfo{volume}{25},
  \bibinfo{number}{2} (\bibinfo{year}{2020}), \bibinfo{pages}{1063--1094}.
\newblock


\bibitem[Bacchelli and Bird(2013)]%
        {bacchelli2013expectations}
\bibfield{author}{\bibinfo{person}{Alberto Bacchelli} {and}
  \bibinfo{person}{Christian Bird}.} \bibinfo{year}{2013}\natexlab{}.
\newblock \showarticletitle{Expectations, outcomes, and challenges of modern
  code review}. In \bibinfo{booktitle}{\emph{Proc.\ IEEE/ACM ICSE'13}}. IEEE,
  \bibinfo{pages}{712--721}.
\newblock


\bibitem[Baum et~al\mbox{.}(2016)]%
        {baum2016factors}
\bibfield{author}{\bibinfo{person}{Tobias Baum}, \bibinfo{person}{Olga Liskin},
  \bibinfo{person}{Kai Niklas}, {and} \bibinfo{person}{Kurt Schneider}.}
  \bibinfo{year}{2016}\natexlab{}.
\newblock \showarticletitle{Factors influencing code review processes in
  industry}. In \bibinfo{booktitle}{\emph{Proc.\ ACM SIGSOFT FSE'16}}.
  \bibinfo{pages}{85--96}.
\newblock


\bibitem[Braz et~al\mbox{.}(2022)]%
        {braz2022less}
\bibfield{author}{\bibinfo{person}{Larissa Braz}, \bibinfo{person}{Christian
  Aeberhard}, \bibinfo{person}{G{\"u}l {\c{C}}alikli}, {and}
  \bibinfo{person}{Alberto Bacchelli}.} \bibinfo{year}{2022}\natexlab{}.
\newblock \showarticletitle{Less is More: Supporting Developers in
  Vulnerability Detection during Code Review}. In
  \bibinfo{booktitle}{\emph{Proc.\ IEEE/ACM ICSE'22}}.
  \bibinfo{pages}{1317--1329}.
\newblock


\bibitem[Bui et~al\mbox{.}(2022)]%
        {bui2022vul4j}
\bibfield{author}{\bibinfo{person}{Quang~Cuong Bui}, \bibinfo{person}{Riccardo
  Scandariato}, {and} \bibinfo{person}{Nicolás E.~Díaz Ferreyra}.}
  \bibinfo{year}{2022}\natexlab{}.
\newblock \showarticletitle{Vul4J: A Dataset of Reproducible Java
  Vulnerabilities Geared Towards the Study of Program Repair Techniques}. In
  \bibinfo{booktitle}{\emph{Proc.\ IEEE/ACM MSR'22}}.
\newblock


\bibitem[Cambronero et~al\mbox{.}(2019)]%
        {cambronero2019characterizing}
\bibfield{author}{\bibinfo{person}{Jos{\'e}~Pablo Cambronero},
  \bibinfo{person}{Jiasi Shen}, \bibinfo{person}{J{\"u}rgen Cito},
  \bibinfo{person}{Elena Glassman}, {and} \bibinfo{person}{Martin Rinard}.}
  \bibinfo{year}{2019}\natexlab{}.
\newblock \showarticletitle{Characterizing developer use of automatically
  generated patches}. In \bibinfo{booktitle}{\emph{Proc.\ VL/HCC'19}}. IEEE,
  \bibinfo{pages}{181--185}.
\newblock


\bibitem[Chi et~al\mbox{.}(2022)]%
        {chi2022seqtrans}
\bibfield{author}{\bibinfo{person}{Jianlei Chi}, \bibinfo{person}{Yu Qu},
  \bibinfo{person}{Ting Liu}, \bibinfo{person}{Qinghua Zheng}, {and}
  \bibinfo{person}{Heng Yin}.} \bibinfo{year}{2022}\natexlab{}.
\newblock \showarticletitle{Seqtrans: Automatic vulnerability fix via sequence
  to sequence learning}.
\newblock \bibinfo{journal}{\emph{IEEE Transactions on Software Engineering}}
  (\bibinfo{year}{2022}).
\newblock


\bibitem[Chong et~al\mbox{.}(2021)]%
        {chong2021assessing:student:code:review}
\bibfield{author}{\bibinfo{person}{Chun~Yong Chong}, \bibinfo{person}{Patanamon
  Thongtanunam}, {and} \bibinfo{person}{Chakkrit Tantithamthavorn}.}
  \bibinfo{year}{2021}\natexlab{}.
\newblock \showarticletitle{Assessing the students' understanding and their
  mistakes in code review checklists: an experience report of 1,791 code review
  checklist questions from 394 students}. In \bibinfo{booktitle}{\emph{Proc.\
  IEEE/ACM ICSE-SEET'21}}. IEEE, \bibinfo{pages}{20--29}.
\newblock


\bibitem[Cohen(2010)]%
        {cohen2010modern}
\bibfield{author}{\bibinfo{person}{Jason Cohen}.}
  \bibinfo{year}{2010}\natexlab{}.
\newblock \showarticletitle{Modern code review}.
\newblock \bibinfo{journal}{\emph{Making Software: What Really Works, and Why
  We Believe It}} (\bibinfo{year}{2010}), \bibinfo{pages}{329--336}.
\newblock


\bibitem[Durieux et~al\mbox{.}(2019)]%
        {durieux2019empirical}
\bibfield{author}{\bibinfo{person}{Thomas Durieux}, \bibinfo{person}{Fernanda
  Madeiral}, \bibinfo{person}{Matias Martinez}, {and} \bibinfo{person}{Rui
  Abreu}.} \bibinfo{year}{2019}\natexlab{}.
\newblock \showarticletitle{Empirical review of Java program repair tools: A
  large-scale experiment on 2,141 bugs and 23,551 repair attempts}. In
  \bibinfo{booktitle}{\emph{Proc.\ ACM ESEC/FSE'19}}.
  \bibinfo{pages}{302--313}.
\newblock


\bibitem[{Food and Drug Administration}(2001)]%
        {food2001guidance}
\bibfield{author}{\bibinfo{person}{{Food and Drug Administration}}.}
  \bibinfo{year}{2001}\natexlab{}.
\newblock \bibinfo{title}{Guidance for industry: Statistical approaches to
  establishing bioequivalence}.
\newblock
\newblock


\bibitem[Fry et~al\mbox{.}(2012)]%
        {fry2012human}
\bibfield{author}{\bibinfo{person}{Zachary~P Fry}, \bibinfo{person}{Bryan
  Landau}, {and} \bibinfo{person}{Westley Weimer}.}
  \bibinfo{year}{2012}\natexlab{}.
\newblock \showarticletitle{A human study of patch maintainability}. In
  \bibinfo{booktitle}{\emph{Proc.\ ACM SIGSOFT ISSTA'12}}.
  \bibinfo{pages}{177--187}.
\newblock


\bibitem[Gon{\c{c}}alves et~al\mbox{.}(2020)]%
        {gonccalves2020explicit}
\bibfield{author}{\bibinfo{person}{Pavl{\'\i}na~Wurzel Gon{\c{c}}alves},
  \bibinfo{person}{Enrico Fregnan}, \bibinfo{person}{Tobias Baum},
  \bibinfo{person}{Kurt Schneider}, {and} \bibinfo{person}{Alberto Bacchelli}.}
  \bibinfo{year}{2020}\natexlab{}.
\newblock \showarticletitle{Do explicit review strategies improve code review
  performance?}. In \bibinfo{booktitle}{\emph{Proc.\ IEEE/ACM MSR'20}}.
  \bibinfo{pages}{606--610}.
\newblock


\bibitem[Gramatica et~al\mbox{.}(2015)]%
        {gramatica2015role}
\bibfield{author}{\bibinfo{person}{Martina~de Gramatica},
  \bibinfo{person}{Katsiaryna Labunets}, \bibinfo{person}{Fabio Massacci},
  \bibinfo{person}{Federica Paci}, {and} \bibinfo{person}{Alessandra
  Tedeschi}.} \bibinfo{year}{2015}\natexlab{}.
\newblock \showarticletitle{The role of catalogues of threats and security
  controls in security risk assessment: an empirical study with ATM
  professionals}. In \bibinfo{booktitle}{\emph{Proc.\ REFSQ'15}}. Springer,
  \bibinfo{pages}{98--114}.
\newblock


\bibitem[Just et~al\mbox{.}(2014)]%
        {just2014defects4j}
\bibfield{author}{\bibinfo{person}{Ren{\'e} Just}, \bibinfo{person}{Darioush
  Jalali}, {and} \bibinfo{person}{Michael~D. Ernst}.}
  \bibinfo{year}{2014}\natexlab{}.
\newblock \showarticletitle{{Defects4J}: A {Database} of existing faults to
  enable controlled testing studies for {Java} programs}. In
  \bibinfo{booktitle}{\emph{Proc.\ ACM SIGSOFT ISSTA'14}}.
  \bibinfo{pages}{437--440}.
\newblock


\bibitem[Kechagia et~al\mbox{.}(2021)]%
        {kechagia2021evaluating}
\bibfield{author}{\bibinfo{person}{Maria Kechagia}, \bibinfo{person}{Sergey
  Mechtaev}, \bibinfo{person}{Federica Sarro}, {and} \bibinfo{person}{Mark
  Harman}.} \bibinfo{year}{2021}\natexlab{}.
\newblock \showarticletitle{Evaluating automatic program repair capabilities to
  repair API misuses}.
\newblock \bibinfo{journal}{\emph{IEEE Transactions on Software Engineering}}
  (\bibinfo{year}{2021}).
\newblock


\bibitem[Kollanus and Koskinen(2009)]%
        {kollanus2009survey}
\bibfield{author}{\bibinfo{person}{Sami Kollanus} {and} \bibinfo{person}{Jussi
  Koskinen}.} \bibinfo{year}{2009}\natexlab{}.
\newblock \showarticletitle{Survey of software inspection research}.
\newblock \bibinfo{journal}{\emph{The Open Software Engineering Journal}}
  \bibinfo{volume}{3}, \bibinfo{number}{1} (\bibinfo{year}{2009}).
\newblock


\bibitem[Lachenbruch(2002)]%
        {lachenbruch2002analysis}
\bibfield{author}{\bibinfo{person}{Peter~A Lachenbruch}.}
  \bibinfo{year}{2002}\natexlab{}.
\newblock \showarticletitle{Analysis of data with excess zeros}.
\newblock \bibinfo{journal}{\emph{Statistical methods in medical research}}
  \bibinfo{volume}{11}, \bibinfo{number}{4} (\bibinfo{year}{2002}),
  \bibinfo{pages}{297--302}.
\newblock


\bibitem[Le~Goues et~al\mbox{.}(2012)]%
        {le2012systematic}
\bibfield{author}{\bibinfo{person}{Claire Le~Goues}, \bibinfo{person}{Michael
  Dewey-Vogt}, \bibinfo{person}{Stephanie Forrest}, {and}
  \bibinfo{person}{Westley Weimer}.} \bibinfo{year}{2012}\natexlab{}.
\newblock \showarticletitle{A systematic study of automated program repair:
  Fixing 55 out of 105 bugs for \$8 each}. In \bibinfo{booktitle}{\emph{Proc.\
  IEEE/ACM ICSE'12}}. \bibinfo{pages}{3--13}.
\newblock


\bibitem[Liu et~al\mbox{.}(2019)]%
        {liu2019tbar}
\bibfield{author}{\bibinfo{person}{Kui Liu}, \bibinfo{person}{Anil Koyuncu},
  \bibinfo{person}{Dongsun Kim}, {and} \bibinfo{person}{Tegawend{\'e}~F
  Bissyand{\'e}}.} \bibinfo{year}{2019}\natexlab{}.
\newblock \showarticletitle{TBar: Revisiting template-based automated program
  repair}. In \bibinfo{booktitle}{\emph{Proc.\ ACM SIGSOFT ISSTA'19}}.
  \bibinfo{pages}{31--42}.
\newblock


\bibitem[Liu et~al\mbox{.}(2021)]%
        {liu2021critical:correcteness}
\bibfield{author}{\bibinfo{person}{Kui Liu}, \bibinfo{person}{Li Li},
  \bibinfo{person}{Anil Koyuncu}, \bibinfo{person}{Dongsun Kim},
  \bibinfo{person}{Zhe Liu}, \bibinfo{person}{Jacques Klein}, {and}
  \bibinfo{person}{Tegawend{\'e}~F Bissyand{\'e}}.}
  \bibinfo{year}{2021}\natexlab{}.
\newblock \showarticletitle{A critical review on the evaluation of automated
  program repair systems}.
\newblock \bibinfo{journal}{\emph{Journal of Systems and Software}}
  \bibinfo{volume}{171} (\bibinfo{year}{2021}), \bibinfo{pages}{110817}.
\newblock


\bibitem[Liu et~al\mbox{.}(2020)]%
        {liu2019icseeval}
\bibfield{author}{\bibinfo{person}{Kui Liu}, \bibinfo{person}{Shangwen Wang},
  \bibinfo{person}{Anil Koyuncu}, \bibinfo{person}{Kisub Kim},
  \bibinfo{person}{Tegawend\'{e}~F. Bissyand\'{e}}, \bibinfo{person}{Dongsun
  Kim}, \bibinfo{person}{Peng Wu}, \bibinfo{person}{Jacques Klein},
  \bibinfo{person}{Xiaoguang Mao}, {and} \bibinfo{person}{Yves~Le Traon}.}
  \bibinfo{year}{2020}\natexlab{}.
\newblock \showarticletitle{On the Efficiency of Test Suite Based Program
  Repair: A Systematic Assessment of 16 Automated Repair Systems for Java
  Programs}. In \bibinfo{booktitle}{\emph{Proc.\ ACM/IEEE ICSE'20}}.
  \bibinfo{pages}{615–627}.
\newblock


\bibitem[Ma et~al\mbox{.}(2017)]%
        {ma2017vurle}
\bibfield{author}{\bibinfo{person}{Siqi Ma}, \bibinfo{person}{Ferdian Thung},
  \bibinfo{person}{David Lo}, \bibinfo{person}{Cong Sun}, {and}
  \bibinfo{person}{Robert~H Deng}.} \bibinfo{year}{2017}\natexlab{}.
\newblock \showarticletitle{Vurle: Automatic vulnerability detection and repair
  by learning from examples}. In \bibinfo{booktitle}{\emph{European Symposium
  on Research in Computer Security}}. Springer, \bibinfo{pages}{229--246}.
\newblock


\bibitem[Martinez et~al\mbox{.}(2017)]%
        {martinez2017automatic:assessment}
\bibfield{author}{\bibinfo{person}{Matias Martinez}, \bibinfo{person}{Thomas
  Durieux}, \bibinfo{person}{Romain Sommerard}, \bibinfo{person}{Jifeng Xuan},
  {and} \bibinfo{person}{Martin Monperrus}.} \bibinfo{year}{2017}\natexlab{}.
\newblock \showarticletitle{Automatic repair of real bugs in java: A
  large-scale experiment on the defects4j dataset}.
\newblock \bibinfo{journal}{\emph{Empirical Software Engineering}}
  \bibinfo{volume}{22}, \bibinfo{number}{4} (\bibinfo{year}{2017}),
  \bibinfo{pages}{1936--1964}.
\newblock


\bibitem[Martinez and Monperrus(2016)]%
        {martinez2016astor}
\bibfield{author}{\bibinfo{person}{Matias Martinez} {and}
  \bibinfo{person}{Martin Monperrus}.} \bibinfo{year}{2016}\natexlab{}.
\newblock \showarticletitle{Astor: A program repair library for java}. In
  \bibinfo{booktitle}{\emph{Proc.\ ACM SIGSOFT ISSTA'16}}.
  \bibinfo{pages}{441--444}.
\newblock


\bibitem[Martinez and Monperrus(2018)]%
        {martinez2018ultra}
\bibfield{author}{\bibinfo{person}{Matias Martinez} {and}
  \bibinfo{person}{Martin Monperrus}.} \bibinfo{year}{2018}\natexlab{}.
\newblock \showarticletitle{Ultra-large repair search space with automatically
  mined templates: The cardumen mode of astor}. In
  \bibinfo{booktitle}{\emph{Proc.\ SSBSE'18}}. \bibinfo{pages}{65--86}.
\newblock


\bibitem[Massacci and Pashchenko(2021a)]%
        {massacci2021technical:magazine}
\bibfield{author}{\bibinfo{person}{Fabio Massacci} {and} \bibinfo{person}{Ivan
  Pashchenko}.} \bibinfo{year}{2021}\natexlab{a}.
\newblock \showarticletitle{Technical Leverage: Dependencies Are a Mixed
  Blessing.}
\newblock \bibinfo{journal}{\emph{IEEE Sec. \& Priv.}} \bibinfo{volume}{19},
  \bibinfo{number}{3} (\bibinfo{year}{2021}), \bibinfo{pages}{58--62}.
\newblock


\bibitem[Massacci and Pashchenko(2021b)]%
        {massacci2021technical:conference}
\bibfield{author}{\bibinfo{person}{Fabio Massacci} {and} \bibinfo{person}{Ivan
  Pashchenko}.} \bibinfo{year}{2021}\natexlab{b}.
\newblock \showarticletitle{Technical leverage in a software ecosystem:
  Development opportunities and security risks}. In
  \bibinfo{booktitle}{\emph{Proc.\ ACM/IEEE ICSE'21}}. IEEE,
  \bibinfo{pages}{1386--1397}.
\newblock


\bibitem[Meyners(2012)]%
        {meyners2012equivalence}
\bibfield{author}{\bibinfo{person}{Michael Meyners}.}
  \bibinfo{year}{2012}\natexlab{}.
\newblock \showarticletitle{Equivalence tests--A review}.
\newblock \bibinfo{journal}{\emph{Food quality and preference}}
  \bibinfo{volume}{26}, \bibinfo{number}{2} (\bibinfo{year}{2012}),
  \bibinfo{pages}{231--245}.
\newblock


\bibitem[Monperrus(2018)]%
        {repair-living-review}
\bibfield{author}{\bibinfo{person}{Martin Monperrus}.}
  \bibinfo{year}{2018}\natexlab{}.
\newblock \bibinfo{booktitle}{\emph{The Living Review on Automated Program
  Repair}}.
\newblock \bibinfo{type}{{T}echnical {R}eport} hal-01956501.
  \bibinfo{institution}{HAL/archives-ouvertes.fr}.
\newblock


\bibitem[Naiakshina et~al\mbox{.}(2017)]%
        {naiakshina2017developers}
\bibfield{author}{\bibinfo{person}{Alena Naiakshina},
  \bibinfo{person}{Anastasia Danilova}, \bibinfo{person}{Christian Tiefenau},
  \bibinfo{person}{Marco Herzog}, \bibinfo{person}{Sergej Dechand}, {and}
  \bibinfo{person}{Matthew Smith}.} \bibinfo{year}{2017}\natexlab{}.
\newblock \showarticletitle{Why do developers get password storage wrong? A
  qualitative usability study}. In \bibinfo{booktitle}{\emph{Proc.\ ACM SIGSAC
  CCS'17}}. \bibinfo{pages}{311--328}.
\newblock


\bibitem[Naiakshina et~al\mbox{.}(2018)]%
        {naiakshina2018deception}
\bibfield{author}{\bibinfo{person}{Alena Naiakshina},
  \bibinfo{person}{Anastasia Danilova}, \bibinfo{person}{Christian Tiefenau},
  {and} \bibinfo{person}{Matthew Smith}.} \bibinfo{year}{2018}\natexlab{}.
\newblock \showarticletitle{Deception task design in developer password
  studies: Exploring a student sample}. In \bibinfo{booktitle}{\emph{Proc.\
  USENIX SOUPS'18}}. \bibinfo{pages}{297--313}.
\newblock


\bibitem[Pashchenko et~al\mbox{.}(2020)]%
        {pashchenko2020qualitative:dependency}
\bibfield{author}{\bibinfo{person}{Ivan Pashchenko}, \bibinfo{person}{Duc-Ly
  Vu}, {and} \bibinfo{person}{Fabio Massacci}.}
  \bibinfo{year}{2020}\natexlab{}.
\newblock \showarticletitle{A qualitative study of dependency management and
  its security implications}. In \bibinfo{booktitle}{\emph{Proc.\ ACM SIGSAC
  CCS'20}}. \bibinfo{pages}{1513--1531}.
\newblock


\bibitem[Pinconschi et~al\mbox{.}(2021)]%
        {pinconschi2021comparative}
\bibfield{author}{\bibinfo{person}{Eduard Pinconschi}, \bibinfo{person}{Rui
  Abreu}, {and} \bibinfo{person}{Pedro Ad{\~a}o}.}
  \bibinfo{year}{2021}\natexlab{}.
\newblock \showarticletitle{A Comparative Study of Automatic Program Repair
  Techniques for Security Vulnerabilities}. In \bibinfo{booktitle}{\emph{Proc.\
  IEEE ISSRE'21}}. IEEE, \bibinfo{pages}{196--207}.
\newblock


\bibitem[Ponta et~al\mbox{.}(2019)]%
        {ponta2019manually}
\bibfield{author}{\bibinfo{person}{Serena~Elisa Ponta}, \bibinfo{person}{Henrik
  Plate}, \bibinfo{person}{Antonino Sabetta}, \bibinfo{person}{Michele Bezzi},
  {and} \bibinfo{person}{C{\'e}dric Dangremont}.}
  \bibinfo{year}{2019}\natexlab{}.
\newblock \showarticletitle{A manually-curated dataset of fixes to
  vulnerabilities of open-source software}. In \bibinfo{booktitle}{\emph{Proc.\
  IEEE/ACM MSR'19}}. \bibinfo{pages}{383--387}.
\newblock


\bibitem[Rigby and Bird(2013)]%
        {rigby2013convergent}
\bibfield{author}{\bibinfo{person}{Peter~C Rigby} {and}
  \bibinfo{person}{Christian Bird}.} \bibinfo{year}{2013}\natexlab{}.
\newblock \showarticletitle{Convergent contemporary software peer review
  practices}. In \bibinfo{booktitle}{\emph{Proc.\ ACM ESEC/FSE'13}}.
  \bibinfo{pages}{202--212}.
\newblock


\bibitem[Rigby et~al\mbox{.}(2014)]%
        {rigby2014peer}
\bibfield{author}{\bibinfo{person}{Peter~C Rigby}, \bibinfo{person}{Daniel~M
  German}, \bibinfo{person}{Laura Cowen}, {and} \bibinfo{person}{Margaret-Anne
  Storey}.} \bibinfo{year}{2014}\natexlab{}.
\newblock \showarticletitle{Peer review on open-source software projects:
  Parameters, statistical models, and theory}.
\newblock \bibinfo{journal}{\emph{ACM Transactions on Software Engineering and
  Methodology (TOSEM)}} \bibinfo{volume}{23}, \bibinfo{number}{4}
  (\bibinfo{year}{2014}), \bibinfo{pages}{1--33}.
\newblock


\bibitem[Rong et~al\mbox{.}(2012)]%
        {rong2012effect}
\bibfield{author}{\bibinfo{person}{Guoping Rong}, \bibinfo{person}{Jingyi Li},
  \bibinfo{person}{Mingjuan Xie}, {and} \bibinfo{person}{Tao Zheng}.}
  \bibinfo{year}{2012}\natexlab{}.
\newblock \showarticletitle{The effect of checklist in code review for
  inexperienced students: An empirical study}. In
  \bibinfo{booktitle}{\emph{Proc.\ IEEE CSEET'12}}. IEEE,
  \bibinfo{pages}{120--124}.
\newblock


\bibitem[Sadowski et~al\mbox{.}(2018)]%
        {sadowski2018modern}
\bibfield{author}{\bibinfo{person}{Caitlin Sadowski}, \bibinfo{person}{Emma
  S{\"o}derberg}, \bibinfo{person}{Luke Church}, \bibinfo{person}{Michal
  Sipko}, {and} \bibinfo{person}{Alberto Bacchelli}.}
  \bibinfo{year}{2018}\natexlab{}.
\newblock \showarticletitle{Modern code review: a case study at {Google}}. In
  \bibinfo{booktitle}{\emph{Proc.\ IEEE/ACM ICSE-SEIP'18}}.
  \bibinfo{pages}{181--190}.
\newblock


\bibitem[Salman et~al\mbox{.}(2015)]%
        {salman2015students}
\bibfield{author}{\bibinfo{person}{Iflaah Salman}, \bibinfo{person}{Ayse~Tosun
  Misirli}, {and} \bibinfo{person}{Natalia Juristo}.}
  \bibinfo{year}{2015}\natexlab{}.
\newblock \showarticletitle{Are students representatives of professionals in
  software engineering experiments?}. In \bibinfo{booktitle}{\emph{Proc.\
  IEEE/ACM ICSE'15}}, Vol.~\bibinfo{volume}{1}. IEEE,
  \bibinfo{pages}{666--676}.
\newblock


\bibitem[Scandariato et~al\mbox{.}(2015)]%
        {scandariato2015descriptive}
\bibfield{author}{\bibinfo{person}{Riccardo Scandariato}, \bibinfo{person}{Kim
  Wuyts}, {and} \bibinfo{person}{Wouter Joosen}.}
  \bibinfo{year}{2015}\natexlab{}.
\newblock \showarticletitle{A descriptive study of Microsoft’s threat
  modeling technique}.
\newblock \bibinfo{journal}{\emph{Requirements Engineering}}
  \bibinfo{volume}{20}, \bibinfo{number}{2} (\bibinfo{year}{2015}),
  \bibinfo{pages}{163--180}.
\newblock


\bibitem[Schuirmann(1981)]%
        {schuirmann1981hypothesis}
\bibfield{author}{\bibinfo{person}{DL Schuirmann}.}
  \bibinfo{year}{1981}\natexlab{}.
\newblock \showarticletitle{On hypothesis-testing to determine if the mean of a
  normal-distribution is contained in a known interval}.
\newblock \bibinfo{journal}{\emph{Biometrics}} \bibinfo{volume}{37},
  \bibinfo{number}{3} (\bibinfo{year}{1981}), \bibinfo{pages}{617--617}.
\newblock


\bibitem[Shahin et~al\mbox{.}(2017)]%
        {shahin2017continuous:review}
\bibfield{author}{\bibinfo{person}{Mojtaba Shahin},
  \bibinfo{person}{Muhammad~Ali Babar}, {and} \bibinfo{person}{Liming Zhu}.}
  \bibinfo{year}{2017}\natexlab{}.
\newblock \showarticletitle{Continuous integration, delivery and deployment: a
  systematic review on approaches, tools, challenges and practices}.
\newblock \bibinfo{journal}{\emph{IEEE Access}}  \bibinfo{volume}{5}
  (\bibinfo{year}{2017}), \bibinfo{pages}{3909--3943}.
\newblock


\bibitem[Tao et~al\mbox{.}(2014)]%
        {tao2014automatically}
\bibfield{author}{\bibinfo{person}{Yida Tao}, \bibinfo{person}{Jindae Kim},
  \bibinfo{person}{Sunghun Kim}, {and} \bibinfo{person}{Chang Xu}.}
  \bibinfo{year}{2014}\natexlab{}.
\newblock \showarticletitle{Automatically generated patches as debugging aids:
  a human study}. In \bibinfo{booktitle}{\emph{Proc.\ ACM SIGSOFT FSE'14}}.
  \bibinfo{pages}{64--74}.
\newblock


\bibitem[Tuma and Scandariato(2018)]%
        {tuma2018two}
\bibfield{author}{\bibinfo{person}{Katja Tuma} {and} \bibinfo{person}{Riccardo
  Scandariato}.} \bibinfo{year}{2018}\natexlab{}.
\newblock \showarticletitle{Two architectural threat analysis techniques
  compared}. In \bibinfo{booktitle}{\emph{European Conference on Software
  Architecture}}. Springer, \bibinfo{pages}{347--363}.
\newblock


\bibitem[Wang et~al\mbox{.}(2020b)]%
        {wang2020automated:correcteness}
\bibfield{author}{\bibinfo{person}{Shangwen Wang}, \bibinfo{person}{Ming Wen},
  \bibinfo{person}{Bo Lin}, \bibinfo{person}{Hongjun Wu},
  \bibinfo{person}{Yihao Qin}, \bibinfo{person}{Deqing Zou},
  \bibinfo{person}{Xiaoguang Mao}, {and} \bibinfo{person}{Hai Jin}.}
  \bibinfo{year}{2020}\natexlab{b}.
\newblock \showarticletitle{Automated patch correctness assessment: How far are
  we?}. In \bibinfo{booktitle}{\emph{Proc.\ IEEE/ACM ASE'20}}.
  \bibinfo{pages}{968--980}.
\newblock


\bibitem[Wang et~al\mbox{.}(2020a)]%
        {wang2020:updates:java}
\bibfield{author}{\bibinfo{person}{Ying Wang}, \bibinfo{person}{Bihuan Chen},
  \bibinfo{person}{Kaifeng Huang}, \bibinfo{person}{Bowen Shi},
  \bibinfo{person}{Congying Xu}, \bibinfo{person}{Xin Peng},
  \bibinfo{person}{Yijian Wu}, {and} \bibinfo{person}{Yang Liu}.}
  \bibinfo{year}{2020}\natexlab{a}.
\newblock \showarticletitle{An empirical study of usages, updates and risks of
  third-party libraries in java projects}. In \bibinfo{booktitle}{\emph{Proc.\
  of ICSME'20}}. IEEE, \bibinfo{pages}{35--45}.
\newblock


\bibitem[Winter et~al\mbox{.}(2022)]%
        {winter2022let}
\bibfield{author}{\bibinfo{person}{Emily~Rowan Winter}, \bibinfo{person}{Vesna
  Nowack}, \bibinfo{person}{David Bowes}, \bibinfo{person}{Steve Counsell},
  \bibinfo{person}{Tracy Hall}, \bibinfo{person}{Saemundur~O Haraldsson}, {and}
  \bibinfo{person}{John Woodward}.} \bibinfo{year}{2022}\natexlab{}.
\newblock \showarticletitle{Let's Talk With Developers, Not About Developers: A
  Review of Automatic Program Repair Research}.
\newblock \bibinfo{journal}{\emph{IEEE Transactions on Software Engineering}}
  (\bibinfo{year}{2022}).
\newblock


\bibitem[Wuyts et~al\mbox{.}(2014)]%
        {wuyts2014empirical}
\bibfield{author}{\bibinfo{person}{Kim Wuyts}, \bibinfo{person}{Riccardo
  Scandariato}, {and} \bibinfo{person}{Wouter Joosen}.}
  \bibinfo{year}{2014}\natexlab{}.
\newblock \showarticletitle{Empirical evaluation of a privacy-focused threat
  modeling methodology}.
\newblock \bibinfo{journal}{\emph{Journal of Systems and Software}}
  \bibinfo{volume}{96} (\bibinfo{year}{2014}), \bibinfo{pages}{122--138}.
\newblock


\bibitem[Ye et~al\mbox{.}(2021)]%
        {ye2021automated:assessment}
\bibfield{author}{\bibinfo{person}{He Ye}, \bibinfo{person}{Matias Martinez},
  {and} \bibinfo{person}{Martin Monperrus}.} \bibinfo{year}{2021}\natexlab{}.
\newblock \showarticletitle{Automated patch assessment for program repair at
  scale}.
\newblock \bibinfo{journal}{\emph{Empirical Software Engineering}}
  \bibinfo{volume}{26}, \bibinfo{number}{2} (\bibinfo{year}{2021}),
  \bibinfo{pages}{1--38}.
\newblock


\bibitem[Yuan and Banzhaf(2018)]%
        {yuan2018arja}
\bibfield{author}{\bibinfo{person}{Yuan Yuan} {and} \bibinfo{person}{Wolfgang
  Banzhaf}.} \bibinfo{year}{2018}\natexlab{}.
\newblock \showarticletitle{Arja: Automated repair of java programs via
  multi-objective genetic programming}.
\newblock \bibinfo{journal}{\emph{IEEE Transactions on software engineering}}
  \bibinfo{volume}{46}, \bibinfo{number}{10} (\bibinfo{year}{2018}),
  \bibinfo{pages}{1040--1067}.
\newblock


\bibitem[Zhang et~al\mbox{.}(2022)]%
        {zhang2022program:correcteness}
\bibfield{author}{\bibinfo{person}{Quanjun Zhang}, \bibinfo{person}{Yuan Zhao},
  \bibinfo{person}{Weisong Sun}, \bibinfo{person}{Chunrong Fang},
  \bibinfo{person}{Ziyuan Wang}, {and} \bibinfo{person}{Lingming Zhang}.}
  \bibinfo{year}{2022}\natexlab{}.
\newblock \showarticletitle{Program Repair: Automated vs. Manual}.
\newblock \bibinfo{journal}{\emph{arXiv preprint arXiv:2203.05166}}
  (\bibinfo{year}{2022}).
\newblock


\bibitem[Zhang et~al\mbox{.}(2021)]%
        {zhang2021data}
\bibfield{author}{\bibinfo{person}{Ying Zhang}, \bibinfo{person}{Mahir Kabir},
  \bibinfo{person}{Ya Xiao}, \bibinfo{person}{Na Meng}, {et~al\mbox{.}}}
  \bibinfo{year}{2021}\natexlab{}.
\newblock \showarticletitle{Data-Driven Vulnerability Detection and Repair in
  Java Code}.
\newblock \bibinfo{journal}{\emph{arXiv preprint arXiv:2102.06994}}
  (\bibinfo{year}{2021}).
\newblock


\end{thebibliography}

\end{document}